\documentclass[aps,twocolumn,amsmath,amssymb,superscriptaddress,prl]{revtex4}
\usepackage{graphicx}

\begin{document} 
\title{
Equation of state of superfluid neutron matter and the calculation of $^1S_0$ pairing gap.
} 

\author{S. Gandolfi} 
\affiliation{S.I.S.S.A., International School of Advanced Studies,
via Beirut 2/4, 34014 Trieste, Italy}
\affiliation{INFN, Sezione di Trieste, Trieste, Italy}
\author{A. Yu. Illarionov} 
\affiliation{S.I.S.S.A., International School of Advanced Studies,
via Beirut 2/4, 34014 Trieste, Italy}
\author{S. Fantoni}
\affiliation{S.I.S.S.A., International School of Advanced Studies,
via Beirut 2/4, 34014 Trieste, Italy}
\affiliation{INFM, CNR--DEMOCRITOS National Supercomputing Center, Trieste, Italy}
\author{F. Pederiva}
\affiliation{Dipartimento di Fisica, University of Trento,
via Sommarive 14, I-38050 Povo, Trento Italy}
\affiliation{CNR--DEMOCRITOS National Supercomputing Center, Trieste, Italy}
\author{K. E. Schmidt}
\affiliation{Department of Physics, Arizona State University, Tempe, AZ, USA}

\begin{abstract}
We present a Quantum Monte Carlo study of the zero temperature equation of state 
of neutron matter  and the computation of the $^1S_0$ pairing gap in the low-density 
regime with $\rho<0.04$ fm$^{-3}$. The system is described by a non-relativistic nuclear
Hamiltonian including both two-- and three--nucleon interactions of the Argonne
and Urbana type. This model interaction provides very accurate 
results in the calculation of the binding energy of light nuclei.
A suppression of the gap with respect to the pure BCS theory is found, but sensibly weaker
than in other  works that attempt to include polarization effects in an approximate way.
\end{abstract}

\maketitle

Superfluidity of neutron  matter is currently a subject of great interest across
the astrophysics, nuclear physics, and many--body physics communities. 
Pairing of nucleons may occur in different channels. At densities relevant for
the phenomenology of the stellar outer core, a crucial role is played by 
the energy gap of the $^1S_0$ paired state. It has direct consequences on the
low density equation of state, the cooling process, the phenomenology 
of glitches\cite{dean03,monrozeau07}, and on neutrino emission. 
In addition, the effect of pairing of
neutrons in the low density regime play an important role also in the 
pairing effects observed in neutron-rich nuclei, where the energy
is sensibly more stable if the total number of nucleons A is 
even\cite{gandolfi06} (see for example Ref. \cite{dean03} for a review).

 In neutron matter superfluidity of the BCS type occurs up to densities
of order $2\times10^{-2}$ fm$^{-3}$, corresponding to a value of the
parameter $\xi=k_Fa \leq -14.8$, where $a$ is the scattering length,
which for a pure neutron matter is large and negative ($a\simeq-18.5$ fm).
Such value of $\xi$ gives rise to a weak BCS superfluid\cite{heiselberg00b},
and, at the same time, induces strong correlations among neutrons.

\begin{figure}[ht]
\vspace{0.5cm}
\begin{center}
\includegraphics[width=7cm]{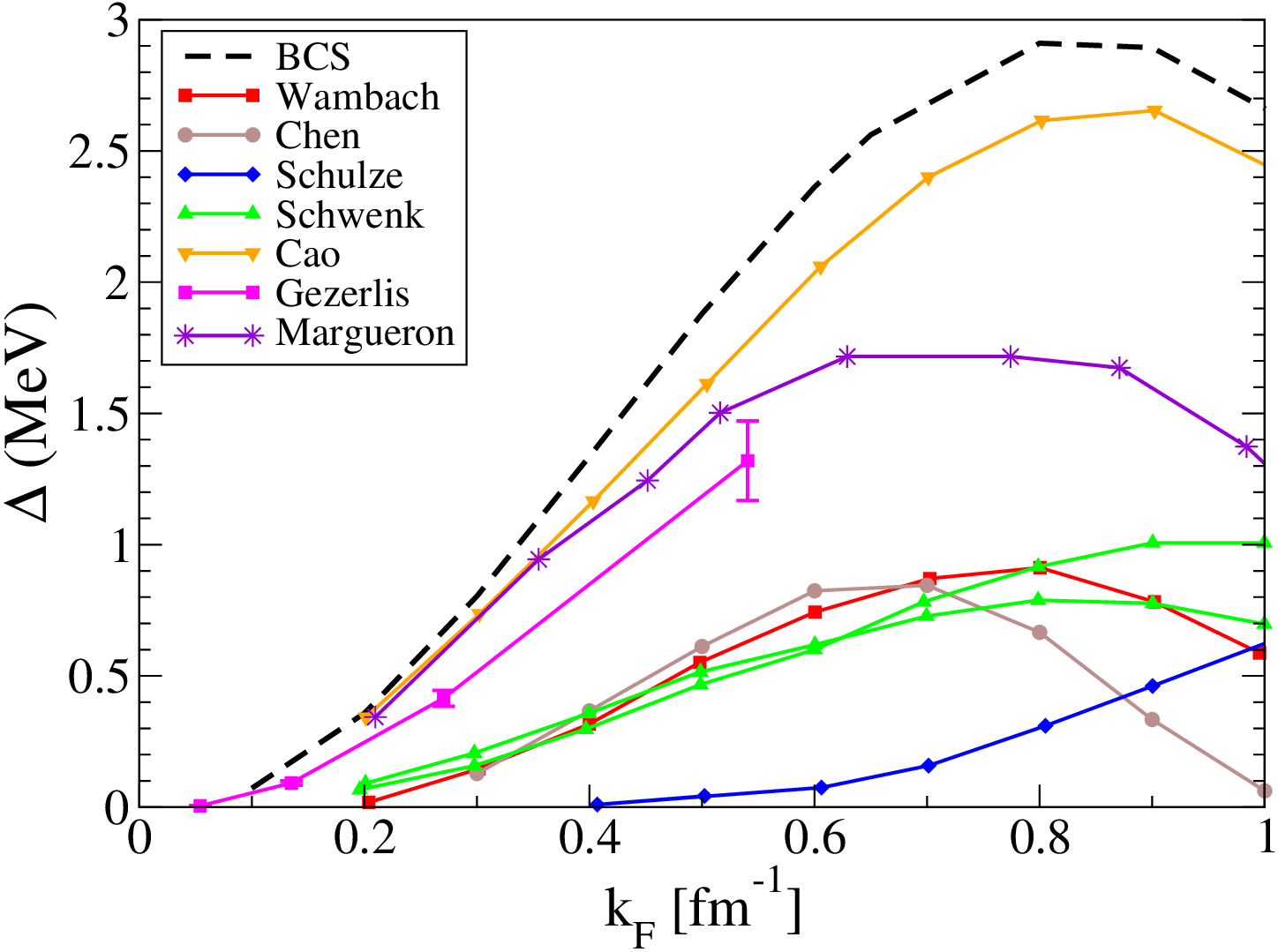}
\caption{(color online) 
The $^1S_0$ pairing gap of neutron matter as a function of the Fermi momentum $k_F$ computed 
with different methods. In the figure we display works of Wambach et al.\cite{wambach93},
Chen et al.\cite{chen93}, Schulze et al.\cite{schulze96}, Schwenk et al.\cite{schwenk03}, Cao et
al.\cite{cao06}, Gezerlis and Carlson\cite{gezerlis08} and Margueron et al.\cite{margueron08}. All the
results are compared with a BCS calculation (dashed line).}
\label{fig:gap1}
\end{center}
\end{figure}

This fact leads  to a second main reason of interest, i.e. the study
of the interplay of the correlations induced by the strong repulsion
among neutrons at short distance, and by the spin dependent forces with
a strong tensor component, and the occurrence of the BCS state.
A clean signature of this effect can be found in the behavior of the
BCS energy gap $\Delta$. This quantity has been computed by several authors
within many different approximation schemes. A mean-field BCS treatment
gives a peak value of the energy gap  $\Delta\simeq 3$ MeV at $k_F\sim0.8$ fm$^{-1}$,
almost irrespective of the NN potential\cite{fabrocini08,hebeler07}. 
This is not surprising because
all NN potentials are fitted to reproduce the same S-- and P--wave
components in the scattering experiments.  The situation becomes more 
intricate when so-called {\it polarization effects}, i.e. the
interaction with the surrounding medium are introduced. There
are two ways to attack the problem. The first is still based on the solution
of the BCS equation(a two body problem) with an effective interaction which
approximates the background effects. Alternatively, a more rigorous calculation
should include many--body effects by solving the many--body problem with
the full interaction, and compute the gap between the BCS and normal state
directly as the energy difference. At this level, an accurate computation
of the BCS gap can be obtained only if 1) A realistic interaction including
all relevant contributions (and in particular hard--core repulsion and tensor)
and 2) An ab-initio computational method is employed. 
The present disagreement among different estimates of the energy gap presently
available, and which is shown in Fig. \ref{fig:gap1}, can be explained by the lack of
at least one of these two conditions.
 
The inclusion of polarization effects shows some general trends in the behavior of the
BCS gap. 
It has been pointed out that the screening by the medium  could strongly reduce the 
pairing strength in the $^1S_0$ channel. Actually, the diverse methods used
in connection with realistic NN interactions all give 
a maximum for $\Delta$ between 0.7 and 1.0 MeV at a density corresponding to
a Fermi momentum between 0.7 and 1.0 fm$^{-1}$\cite{schulze96,schwenk03}. 
Recent Brueckner  theory calculations \cite{cao06} and Hartree Fock
calculations\cite{margueron08} give a remarkably larger value ($\sim$ 2.5 MeV).
A recent quantum Monte Carlo (QMC) calculation\cite{gezerlis08}, 
using the AV18 NN potential projected in the $^1S_0$ channel,
gave in the very low density range corresponding  to $k_F\leq$0.55 fm$^{-1}$
a pairing gap sensibly larger  with respect to previous results, yet lower than the 
Brueckner calculation of Ref. \cite{cao06}, and Hartree Fock calculations
of Ref. \cite{margueron08}.

In this letter we propose a systematic computation 
of the $^1S_0$ pairing gap in 
neutron matter as a function of the density of the system
in which both the conditions of using
a complete realistic interaction (an Argonne-Urbana class potential) and an accurate
ab-initio method (the Auxiliary Field Diffusion Monte Carlo
within the fixed phase approximation) are fullfilled, with
the aim of benchmarking existing results. 
QMC techniques have the advantage of accurately solving for the many-body
ground state, and provide a powerful tool to study a wide range
of systems both finite and homogeneous.
The Auxiliary Field Diffusion Monte Carlo (AFDMC) is particularly well suited
to deal with large nucleonic systems\cite{schmidt99}. In particular,
the AFDMC gives accurate results for the
properties of nuclei\cite{gandolfi07b}, symmetric nuclear matter\cite{gandolfi07}
and neutron matter\cite{sarsa03}, for which studies performed with Green's
function Monte Carlo are limited to 14 neutrons in a periodic box\cite{carlson03}.
The efficiency of AFDMC lies in the fact that spin states of nucleons
are sampled, rather then explicitly summed, by means of a
Hubbard-Stratonovich transformation. It can be extended to systems
with over a hundred nucleons. Larger systems are needed to
rule out finite size effects in the estimates. Moreover, we were recently
able to overcome some technical issues related to the approximation used
to cope with the Fermion sign problem by turning from the constrained path
approximation to a fixed phase approximation. This change resolved
several issues related to discrepancies with existing Green's Function
Monte Carlo calculations\cite{gandolfi07,gandolfi07b,gandolfi07c,gandolfi08x}.
One of the consequences of this change is a major improvement in the
accuracy of the AFDMC calculations for low density neutron matter and
a corresponding improvement in the accuracy of the calculated energy gaps
over previous AFDMC results\cite{fabrocini05}.

Our calculations  for bulk  neutron matter are based on a non-relativistic
Hamiltonian of $N$ neutrons in a periodic box:
\begin{equation}
H=-\frac{\hbar^2}{2m}\sum_{i=1}^N\nabla_i^2+\sum_{i<j}v_{ij}+\sum_{i<j<k}V_{ijk} \,,
\end{equation}
where $i$ and $j$ are neutron indices, and $v_{ij}$ and $V_{ijk}$ are respectively
the two-- and the three-nucleon interactions (namely the Argonne
$v_8'$ (AV8') and
the Urbana IX (UIX). The AV8'\cite{wiringa02} interactions is a simpler 
form of the more accurate Argonne $v_{18}$ (AV18)\cite{wiringa95};
it contains only
8 operators instead of 18, it preserves the same isoscalar part of AV18
in S and P partial waves as well as in the $^3D_1$ wave and its 
coupling to $^3S_1$, and it correctly gives the experimental deuteron energy. 
The advantage of using AV8' rather then AV18 is that it has a more
suitable form for the AFDMC calculation. 
In the low density regime where $\rho<0.04$ fm$^{-3}$ the AV8' gives the same energy
as AV18 within 3\%\cite{gandolfi08x}.
The Urbana UIX was fitted to correct the overbinding of AV18 in
the ground state of light nuclei and to  reproduce the empirical value
of the equilibrium density of nuclear matter\cite{pudliner95}.

We consider the {\it full} nuclear Hamiltonian, instead of projecting
in the pairing channel only, in order to include {\it all} the many-body
correlations in the system that the effective bare interactions eventually miss.

The AFDMC calculations start from a Jastrow-BCS trial wave function of the form
\begin{equation}
\psi_T=[\prod_{i<j}f_J(r_{ij})] \Phi_{BCS}(\mathbf R,S) \,,
\end{equation}
where $R=\{\mathbf r_1...\mathbf r_N\}$ and $S=\{s_1...s_N\}$ are the
space and spin coordinates of the neutrons.
The factor $f_J$ is the central component of the Jastrow correlation computed
in the FHNC/SOC scheme. 
Its only
role is to avoid the overlap between neutrons, and its detailed form has no
influence on the final result.
The $\Phi_{BCS}$ antisymmetric function is built as a Pfaffian\cite{bajdich06b}
of both paired and single particle orbitals. Paired orbitals are defined by
\begin{equation}
\phi(\mathbf r_{ij},s_i,s_j)=\sum_\alpha\frac{v_{k_\alpha}}{u_{k_\alpha}}
e^{i \mathbf k_\alpha\cdot\mathbf r_{ij}}\chi(s_i,s_j) \,,
\end{equation}
where $\chi$ is a spin function coupling two neutrons in the
singlet state. The single particle orbitals are plane waves fitting 
Born-von Karman periodic boundary conditions. The coefficients
$u$ and $v$ entering the paired orbitals are provided by a Correlated 
Basis Functions (CBF) calculation\cite{fabrocini08}. In the case of even $N$,
no single particle wave functions are considered in the Pfaffian, while if
$N$ is odd  the single particle plane wave accommodating the unpaired
neutron is chosen in order to minimize the energy of the system.
Finite-size effects due 
to the truncation of the potential are reduced following the common
procedures described in\cite{sarsa03}. 

\begin{figure}[t!]
\vspace{0.5cm}
\begin{center}
\includegraphics[width=7cm]{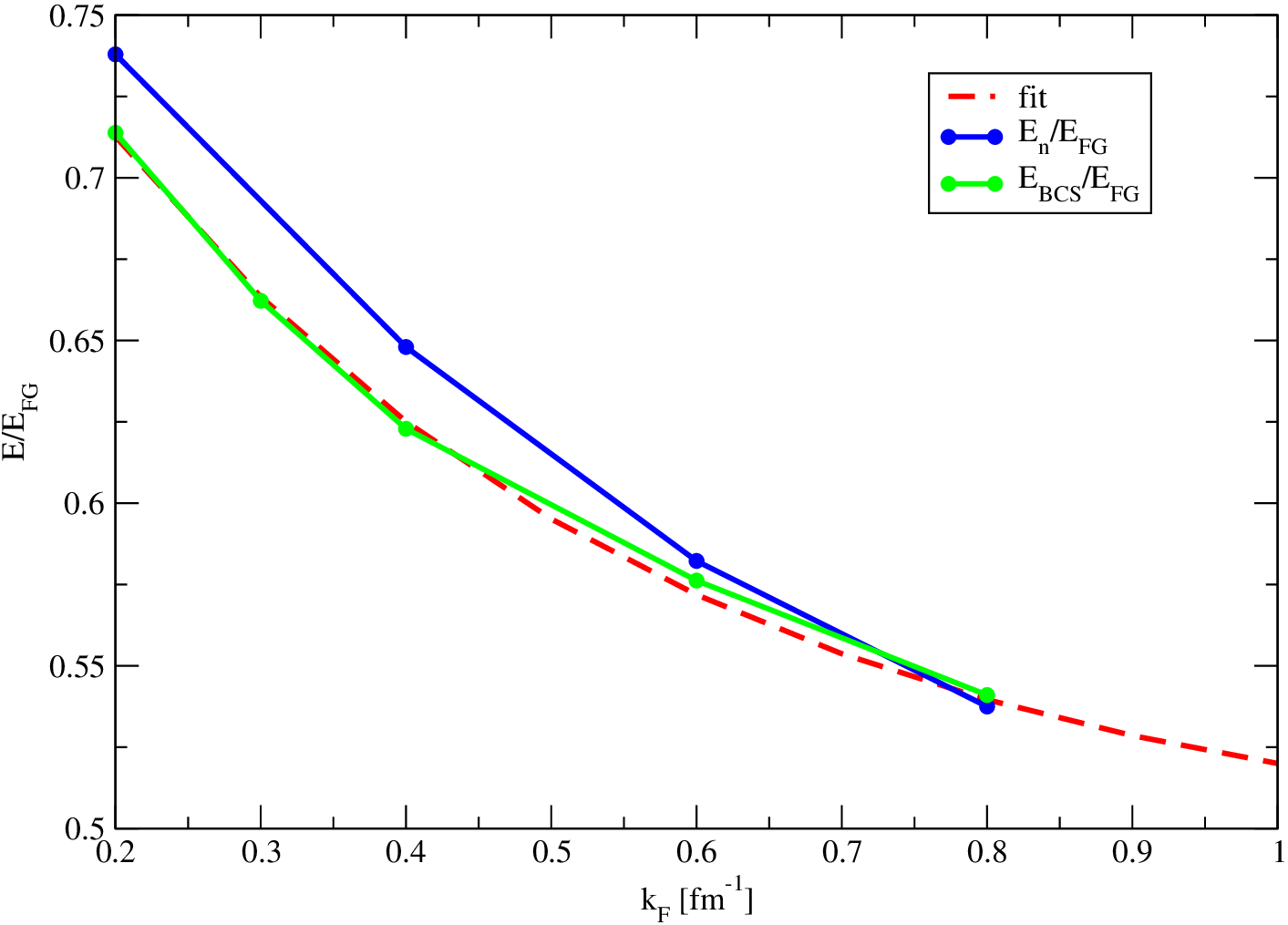}
\caption{(color online) The EOS of neutron matter in the low-density regime. The two calculations were performed 
using different trial wave functions modeling a normal and a BCS state. The fit is a guide to the eye.}
\label{fig:eos}
\end{center}
\end{figure}

Because AFDMC projects out the lowest energy state with the same symmetry
and phase of the trial wave function from which the projection is started, once the
character of the initial
state (BCS or normal Fermi liquid) is given, the computed energy will refer to
that particular phase. It is therefore possible to compare the two equations
of state, and discuss the relative stability. In Fig. \ref{fig:eos} we display the 
resulting values as a function of $k_F$ normalized to the corresponding
Fermi gas energy. In almost all of the range of
densities considered the BCS phase is stable with respect to
the normal Fermi liquid, although the relative energy difference never exceeds 4\%.


When the Fermi wavevector increases beyond $k_F$=0.6 fm$^{-1}$ the normal
state is energetically more favorite than the BCS state,
with an energy difference smaller than 1\% of the total energy.
We can therefore conclude that in the low-density regime neutron matter
is in a $^1S_0$ superfluid phase. In this regime the neutron-neutron 
interaction is dominated by this channel, having a scattering length
of about $a$=--18.5 fm.
Quantum Monte Carlo calculations of the EOS of dilute cold Fermions showed
that in the unitary limit (when $ak_F\rightarrow-\infty$), the ratio between
the energy of the system and the energy of the Fermi gas is 
$\xi$=0.42(1)\cite{carlson03,astrakharchik04,carlson05}.
The deviation from this asymptotic value is a measure of the relevance of the 
details of the interaction in determining the equation of state of the system in the range
of densities considered.  One should also consider the fact that 
at larger densities  the effect of pairing in scattering channels
other than the $^1S_0$ becomes important.

In a full many--body calculation the superfluid gap can be evaluated by 
using the difference:
\begin{equation}
\label{eq:gap}
\Delta(N)=E(N)-\frac{1}{2}[E(N+1)+E(N-1)] \,.
\end{equation}
It should be noted that the above expression is valid only if $E(N)$, $E(N+1)$ and $E(N-1)$
are computed by keeping the volume $V$ of the system fixed. This means that
the density would be different in the $N$, $N+1$, $N-1$ neutrons systems. 
Because our simulations are usually performed at fixed density, we checked the
dependence of the energy on the constraint used.
Considering a  number of particles around $N=14$, which is the lowest number of
neutrons used in the simulations, and therefore the worst case scenario, we  
evaluated the gap at fixed volume first, and then at fixed density.
The difference in the results is well within statistical errors. 

Several simulations at different values of $N$ were performed
in order to evaluate the gap and the  corresponding statistical error bars.
A first check concerned the dependence of the gap estimate on the 
number of neutron used in evaluating the difference in Eq.(\ref{eq:gap}).
The values of Fermi momentum considered for the check were 0.4, 0.6, and 0.8 fm$^{-1}$,
and the numbers of neutrons were taken in the ranges $N$=12$\div$18 and N=62$\div$68.
For each case, we evaluated the gap around the odd $N$ according to Eq. \ref{eq:gap}. 
At each density $\Delta$(66) (the averaged gap between $N$=62 and 68) is always smaller
than $\Delta$(14). 
The same behavior was also observed in the QMC calculation of Gezerlis and Carlson\cite{gezerlis08}
using the simple interaction projected in the pairing channel.
In that paper computations were extended also to $N$=92. The gap values
 $\Delta$(66) and  $\Delta$(92) are equal within error bars and 
approach the infinite limit in the same way as in the mean field BCS calculation. 
Unfortunately in QMC simulations, in the absence of a correlated
sampling scheme, it is impossible to use arbitrarily large values
of $N$, because the gap has to be evaluated as the difference among total
energies. This means that the accuracy required in the evaluation of the
energy makes the computational time increase with $\sqrt{N}$,
in addition to the $N^3$ standard scaling for Fermion simulations.

A finite size effect might be connected to the relative
size of the neutrons' Cooper pair and the simulation box.
The Heisenberg uncertainty principle can be used to estimate the dimension of a
Cooper pair as
\[
\delta x \simeq \frac{\epsilon_F}{\Delta\cdot k_F}
\] 
where $\epsilon_F$ is the Fermi energy per particle. Taking $\Delta$=2 MeV,
and $k_F$=0.8 fm$^{-1}$, we have therefore $\delta x \simeq $ 6.6 fm, 
much smaller than the typical box size, which for 66 neutrons is $\sim$16 fm. A consequence of this analysis is also
that relevant correlation lengths should be all contained in the simulation
box, implying that an explicit inclusion in the wave function of long range effects (which
are automatically included in mean field calculations) should not lead to 
significant differences in the results.

\begin{figure}[ht]
\vspace{0.5cm}
\begin{center}
\includegraphics[width=7cm]{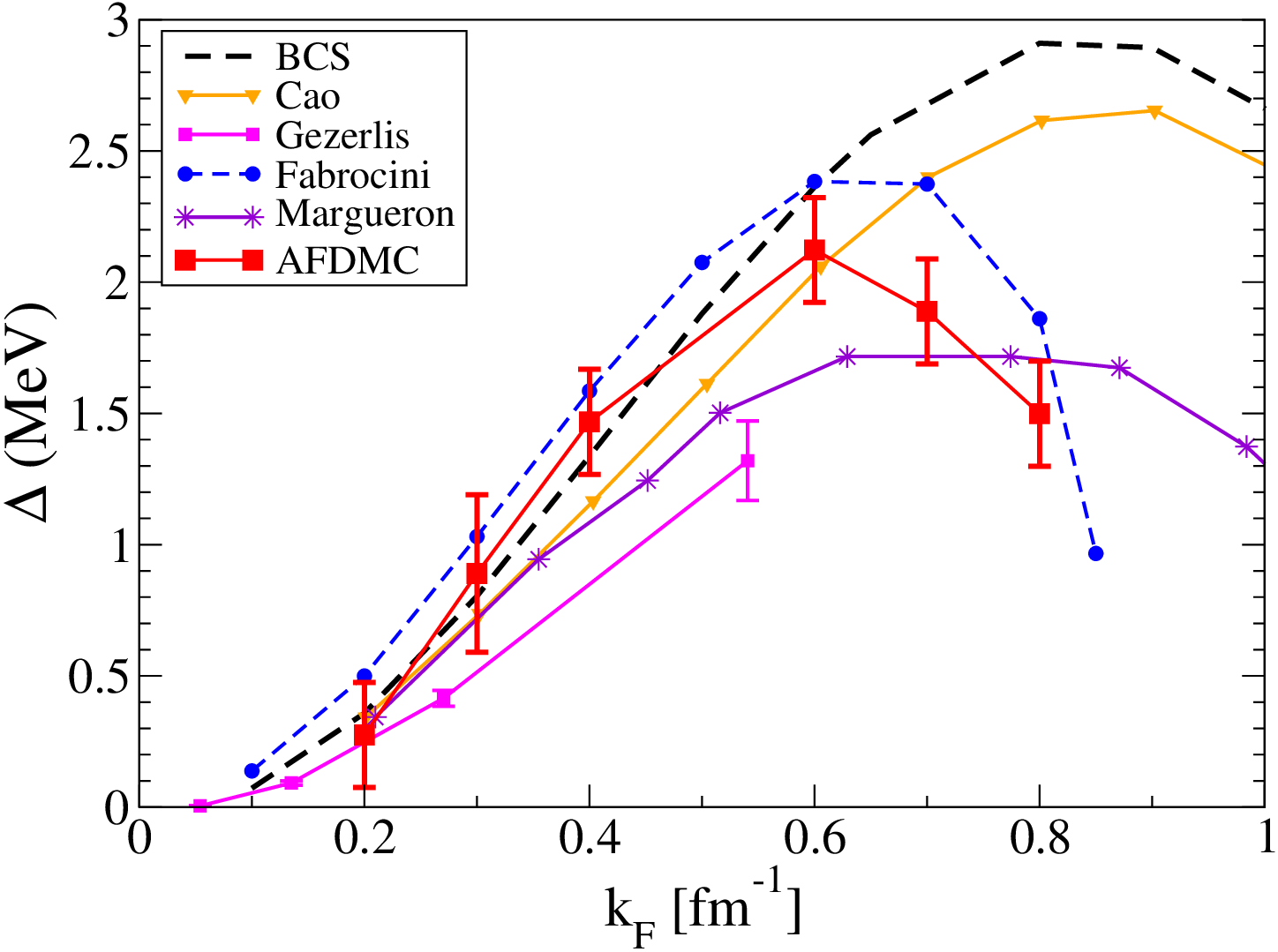}
\caption{(color online) AFDMC calculation of the $^1S_0$ pairing gap of neutron matter as a 
function of the Fermi momentum $k_F$ and compared with more recent results. The AFDMC results 
are indicated by red points with the statistical error bars. Other results are some of 
those displayed in Fig. \ref{fig:gap1}, and the blue dashed line is the CBF calculation
of Fabrocini et al.\cite{fabrocini08}.}
\label{fig:gap2}
\end{center}
\end{figure}
We report in Fig. \ref{fig:gap2} the estimate of $^1S_0$ superfluid gap
as a function of the Fermi momentum $k_F$. The AFDMC points are compared with
results of the CBF calculation\cite{fabrocini08} used to determine the BCS
coefficients entering the trial wave function. We also display, for sake
of comparison, the family of more recent calculations.
As can be seen, the AFDMC calculations give values of $\Delta$  lower 
than those of the CBF calculation. This behavior is opposite to that reported
in a previous paper in which the same comparison was made\cite{fabrocini05}.
The difference can be attributed both to the larger number of neutrons used
in the present work, and to the use of the fixed phase approximation instead of
the constrained path approximation to keep the sign problem under control.

We confirm the depletion of the superfluid gap with respect the BCS result.
However, our results are leaning towards the calculations giving 
a maximum value of the gap of order 2 MeV.
The other available QMC calculations by Gezerlis and Carlson\cite{gezerlis08} differ
within errorbars  for densities corresponding to $k_F<$0.3 fm$^{-1}$.
At $k_F$=0.55 fm$^{-1}$ they predict a gap about 30\% smaller with respect 
to the AFDMC estimate. We believe that such difference comes from the fact
that when increasing the density, the correlations induced by the interactions
in channels others than $^1S_0$ become more and more important, and give
a sizeable contribution to the value of the energy, and consequently to the gap.

We thank J. Carlson, J. Margueron and C. Pethick for useful discussions.
Calculations were partially performed on the BEN cluster at ECT* in Trento, under a grant
for supercomputing projects, and partially on the HPC facility "WIGLAF" of the Department of Physics, 
University of Trento.
This work was supported in part by NSF grants PHY-0456609 and PHY-0757703.

\bibliographystyle{apsrev}
\bibliography{biblio}

\begin{thebibliography}{27}
\expandafter\ifx\csname natexlab\endcsname\relax\def\natexlab#1{#1}\fi
\expandafter\ifx\csname bibnamefont\endcsname\relax
  \def\bibnamefont#1{#1}\fi
\expandafter\ifx\csname bibfnamefont\endcsname\relax
  \def\bibfnamefont#1{#1}\fi
\expandafter\ifx\csname citenamefont\endcsname\relax
  \def\citenamefont#1{#1}\fi
\expandafter\ifx\csname url\endcsname\relax
  \def\url#1{\texttt{#1}}\fi
\expandafter\ifx\csname urlprefix\endcsname\relax\def\urlprefix{URL }\fi
\providecommand{\bibinfo}[2]{#2}
\providecommand{\eprint}[2][]{\url{#2}}

\bibitem[{\citenamefont{Dean and Hjorth-Jensen}(2003)}]{dean03}
\bibinfo{author}{\bibfnamefont{D.~J.} \bibnamefont{Dean}} \bibnamefont{and}
  \bibinfo{author}{\bibfnamefont{M.}~\bibnamefont{Hjorth-Jensen}},
  \bibinfo{journal}{Rev. Mod. Phys.} \textbf{\bibinfo{volume}{75}},
  \bibinfo{pages}{607} (\bibinfo{year}{2003}).

\bibitem[{\citenamefont{Monrozeau et~al.}(2007)\citenamefont{Monrozeau,
  Margueron, and Sandulescu}}]{monrozeau07}
\bibinfo{author}{\bibfnamefont{C.}~\bibnamefont{Monrozeau}},
  \bibinfo{author}{\bibfnamefont{J.}~\bibnamefont{Margueron}},
  \bibnamefont{and}
  \bibinfo{author}{\bibfnamefont{N.}~\bibnamefont{Sandulescu}},
  \bibinfo{journal}{Phys. Rev. C} \textbf{\bibinfo{volume}{75}},
  \bibinfo{pages}{065807} (\bibinfo{year}{2007}).

\bibitem[{\citenamefont{Gandolfi et~al.}(2006)\citenamefont{Gandolfi, Pederiva,
  Fantoni, and Schmidt}}]{gandolfi06}
\bibinfo{author}{\bibfnamefont{S.}~\bibnamefont{Gandolfi}},
  \bibinfo{author}{\bibfnamefont{F.}~\bibnamefont{Pederiva}},
  \bibinfo{author}{\bibfnamefont{S.}~\bibnamefont{Fantoni}}, \bibnamefont{and}
  \bibinfo{author}{\bibfnamefont{K.~E.} \bibnamefont{Schmidt}},
  \bibinfo{journal}{Phys. Rev. C} \textbf{\bibinfo{volume}{73}},
  \bibinfo{pages}{044304} (\bibinfo{year}{2006}).

\bibitem[{\citenamefont{Heiselberg and Hjorth-Jensen}(2000)}]{heiselberg00b}
\bibinfo{author}{\bibfnamefont{H.}~\bibnamefont{Heiselberg}} \bibnamefont{and}
  \bibinfo{author}{\bibfnamefont{M.}~\bibnamefont{Hjorth-Jensen}},
  \bibinfo{journal}{Phys. Rep.} \textbf{\bibinfo{volume}{328}},
  \bibinfo{pages}{237} (\bibinfo{year}{2000}).

\bibitem[{\citenamefont{Fabrocini et~al.}(2008)\citenamefont{Fabrocini,
  Fantoni, Illarionov, and Schmidt}}]{fabrocini08}
\bibinfo{author}{\bibfnamefont{A.}~\bibnamefont{Fabrocini}},
  \bibinfo{author}{\bibfnamefont{S.}~\bibnamefont{Fantoni}},
  \bibinfo{author}{\bibfnamefont{A.~Y.} \bibnamefont{Illarionov}},
  \bibnamefont{and} \bibinfo{author}{\bibfnamefont{K.~E.}
  \bibnamefont{Schmidt}}, \bibinfo{journal}{Nucl. Phys. A}
  \textbf{\bibinfo{volume}{803}}, \bibinfo{pages}{137} (\bibinfo{year}{2008}).

\bibitem[{\citenamefont{Hebeler et~al.}(2007)\citenamefont{Hebeler, Schwenk,
  and Friman}}]{hebeler07}
\bibinfo{author}{\bibfnamefont{K.}~\bibnamefont{Hebeler}},
  \bibinfo{author}{\bibfnamefont{A.}~\bibnamefont{Schwenk}}, \bibnamefont{and}
  \bibinfo{author}{\bibfnamefont{B.}~\bibnamefont{Friman}},
  \bibinfo{journal}{Phys. Lett. B} \textbf{\bibinfo{volume}{648}},
  \bibinfo{pages}{176} (\bibinfo{year}{2007}).

\bibitem[{\citenamefont{J.~Wambach and Pines}(1993)}]{wambach93}
\bibinfo{author}{\bibfnamefont{T.~L.~A.} \bibnamefont{J.~Wambach}}
  \bibnamefont{and} \bibinfo{author}{\bibfnamefont{D.}~\bibnamefont{Pines}},
  \bibinfo{journal}{Nucl. Phys. A} \textbf{\bibinfo{volume}{555}},
  \bibinfo{pages}{128} (\bibinfo{year}{1993}).

\bibitem[{\citenamefont{Chen et~al.}(1993)\citenamefont{Chen, Clark, Dav\'e,
  and Khodel}}]{chen93}
\bibinfo{author}{\bibfnamefont{J.~M.~C.} \bibnamefont{Chen}},
  \bibinfo{author}{\bibfnamefont{J.~W.} \bibnamefont{Clark}},
  \bibinfo{author}{\bibfnamefont{R.~D.} \bibnamefont{Dav\'e}},
  \bibnamefont{and} \bibinfo{author}{\bibfnamefont{V.~V.}
  \bibnamefont{Khodel}}, \bibinfo{journal}{Nucl. Phys. A}
  \textbf{\bibinfo{volume}{555}}, \bibinfo{pages}{59} (\bibinfo{year}{1993}).

\bibitem[{\citenamefont{Schulze et~al.}(1996)\citenamefont{Schulze, Cugnon,
  Lejeune, Baldo, and Lombardo}}]{schulze96}
\bibinfo{author}{\bibfnamefont{H.~J.} \bibnamefont{Schulze}},
  \bibinfo{author}{\bibfnamefont{J.}~\bibnamefont{Cugnon}},
  \bibinfo{author}{\bibfnamefont{A.}~\bibnamefont{Lejeune}},
  \bibinfo{author}{\bibfnamefont{M.}~\bibnamefont{Baldo}}, \bibnamefont{and}
  \bibinfo{author}{\bibfnamefont{U.}~\bibnamefont{Lombardo}},
  \bibinfo{journal}{Phys. Lett. B} \textbf{\bibinfo{volume}{375}},
  \bibinfo{pages}{1} (\bibinfo{year}{1996}).

\bibitem[{\citenamefont{Schwenk et~al.}(2003)\citenamefont{Schwenk, Friman, and
  Brown}}]{schwenk03}
\bibinfo{author}{\bibfnamefont{A.}~\bibnamefont{Schwenk}},
  \bibinfo{author}{\bibfnamefont{B.}~\bibnamefont{Friman}}, \bibnamefont{and}
  \bibinfo{author}{\bibfnamefont{G.~E.} \bibnamefont{Brown}},
  \bibinfo{journal}{Nucl. Phys. A} \textbf{\bibinfo{volume}{713}},
  \bibinfo{pages}{191} (\bibinfo{year}{2003}).

\bibitem[{\citenamefont{Cao et~al.}(2006)\citenamefont{Cao, Lombardo, and
  Schuck}}]{cao06}
\bibinfo{author}{\bibfnamefont{L.~G.} \bibnamefont{Cao}},
  \bibinfo{author}{\bibfnamefont{U.}~\bibnamefont{Lombardo}}, \bibnamefont{and}
  \bibinfo{author}{\bibfnamefont{P.}~\bibnamefont{Schuck}},
  \bibinfo{journal}{Phys. Rev. C} \textbf{\bibinfo{volume}{74}},
  \bibinfo{pages}{064301} (\bibinfo{year}{2006}).

\bibitem[{\citenamefont{Gezerlis and Carlson}(2008)}]{gezerlis08}
\bibinfo{author}{\bibfnamefont{A.}~\bibnamefont{Gezerlis}} \bibnamefont{and}
  \bibinfo{author}{\bibfnamefont{J.}~\bibnamefont{Carlson}},
  \bibinfo{journal}{Phys. Rev. C} \textbf{\bibinfo{volume}{77}},
  \bibinfo{pages}{032801(R)} (\bibinfo{year}{2008}).

\bibitem[{\citenamefont{Margueron et~al.}(2007)\citenamefont{Margueron, Sagawa,
  and Hagino}}]{margueron08}
\bibinfo{author}{\bibfnamefont{J.}~\bibnamefont{Margueron}},
  \bibinfo{author}{\bibfnamefont{H.}~\bibnamefont{Sagawa}}, \bibnamefont{and}
  \bibinfo{author}{\bibfnamefont{K.}~\bibnamefont{Hagino}}
  (\bibinfo{year}{2007}), \eprint{arXiv:712.3644 [nucl-th]}.

\bibitem[{\citenamefont{Schmidt and Fantoni}(1999)}]{schmidt99}
\bibinfo{author}{\bibfnamefont{K.~E.} \bibnamefont{Schmidt}} \bibnamefont{and}
  \bibinfo{author}{\bibfnamefont{S.}~\bibnamefont{Fantoni}},
  \bibinfo{journal}{Phys. Lett. B} \textbf{\bibinfo{volume}{446}},
  \bibinfo{pages}{99} (\bibinfo{year}{1999}).

\bibitem[{\citenamefont{Gandolfi
  et~al.}(2007{\natexlab{a}})\citenamefont{Gandolfi, Pederiva, Fantoni, and
  Schmidt}}]{gandolfi07b}
\bibinfo{author}{\bibfnamefont{S.}~\bibnamefont{Gandolfi}},
  \bibinfo{author}{\bibfnamefont{F.}~\bibnamefont{Pederiva}},
  \bibinfo{author}{\bibfnamefont{S.}~\bibnamefont{Fantoni}}, \bibnamefont{and}
  \bibinfo{author}{\bibfnamefont{K.~E.} \bibnamefont{Schmidt}},
  \bibinfo{journal}{Phys. Rev. Lett.} \textbf{\bibinfo{volume}{99}},
  \bibinfo{pages}{022507} (\bibinfo{year}{2007}{\natexlab{a}}).

\bibitem[{\citenamefont{Gandolfi
  et~al.}(2007{\natexlab{b}})\citenamefont{Gandolfi, Pederiva, Fantoni, and
  Schmidt}}]{gandolfi07}
\bibinfo{author}{\bibfnamefont{S.}~\bibnamefont{Gandolfi}},
  \bibinfo{author}{\bibfnamefont{F.}~\bibnamefont{Pederiva}},
  \bibinfo{author}{\bibfnamefont{S.}~\bibnamefont{Fantoni}}, \bibnamefont{and}
  \bibinfo{author}{\bibfnamefont{K.~E.} \bibnamefont{Schmidt}},
  \bibinfo{journal}{Phys. Rev. Lett.} \textbf{\bibinfo{volume}{98}},
  \bibinfo{pages}{102503} (\bibinfo{year}{2007}{\natexlab{b}}).

\bibitem[{\citenamefont{Sarsa et~al.}(2003)\citenamefont{Sarsa, Fantoni,
  Schmidt, and Pederiva}}]{sarsa03}
\bibinfo{author}{\bibfnamefont{A.}~\bibnamefont{Sarsa}},
  \bibinfo{author}{\bibfnamefont{S.}~\bibnamefont{Fantoni}},
  \bibinfo{author}{\bibfnamefont{K.~E.} \bibnamefont{Schmidt}},
  \bibnamefont{and} \bibinfo{author}{\bibfnamefont{F.}~\bibnamefont{Pederiva}},
  \bibinfo{journal}{Phys. Rev. C} \textbf{\bibinfo{volume}{68}},
  \bibinfo{pages}{024308} (\bibinfo{year}{2003}).

\bibitem[{\citenamefont{Carlson et~al.}(2003)\citenamefont{Carlson, Morales,
  Pandharipande, and Ravenhall}}]{carlson03}
\bibinfo{author}{\bibfnamefont{J.}~\bibnamefont{Carlson}},
  \bibinfo{author}{\bibfnamefont{J.}~\bibnamefont{Morales}},
  \bibinfo{author}{\bibfnamefont{V.~R.} \bibnamefont{Pandharipande}},
  \bibnamefont{and} \bibinfo{author}{\bibfnamefont{D.~G.}
  \bibnamefont{Ravenhall}}, \bibinfo{journal}{Phys. Rev. C}
  \textbf{\bibinfo{volume}{68}}, \bibinfo{pages}{025802}
  (\bibinfo{year}{2003}).

\bibitem[{\citenamefont{Gandolfi}(2007)}]{gandolfi07c}
\bibinfo{author}{\bibfnamefont{S.}~\bibnamefont{Gandolfi}},
  \emph{\bibinfo{title}{The Auxiliary Field Diffusion Monte Carlo Method for
  Nuclear Physics and Nuclear Astrophysics}} (\bibinfo{year}{2007}),
  \bibinfo{note}{{P}h.{D}. thesis}, \eprint{arXiv:0712.1364 [nucl-th]}.

\bibitem[{\citenamefont{Gandolfi}(2008)}]{gandolfi08x}
\bibinfo{author}{\bibfnamefont{S.}~\bibnamefont{Gandolfi}},
  \bibinfo{journal}{private communication}  (\bibinfo{year}{2008}).

\bibitem[{\citenamefont{Fabrocini et~al.}(2005)\citenamefont{Fabrocini,
  Fantoni, Illarionov, and Schmidt}}]{fabrocini05}
\bibinfo{author}{\bibfnamefont{A.}~\bibnamefont{Fabrocini}},
  \bibinfo{author}{\bibfnamefont{S.}~\bibnamefont{Fantoni}},
  \bibinfo{author}{\bibfnamefont{A.~Y.} \bibnamefont{Illarionov}},
  \bibnamefont{and} \bibinfo{author}{\bibfnamefont{K.~E.}
  \bibnamefont{Schmidt}}, \bibinfo{journal}{Phys. Rev. Lett.}
  \textbf{\bibinfo{volume}{95}}, \bibinfo{pages}{192501}
  (\bibinfo{year}{2005}).

\bibitem[{\citenamefont{Wiringa and Pieper}(2002)}]{wiringa02}
\bibinfo{author}{\bibfnamefont{R.~B.} \bibnamefont{Wiringa}} \bibnamefont{and}
  \bibinfo{author}{\bibfnamefont{S.~C.} \bibnamefont{Pieper}},
  \bibinfo{journal}{Phys. Rev. Lett.} \textbf{\bibinfo{volume}{89}},
  \bibinfo{pages}{182501} (\bibinfo{year}{2002}).

\bibitem[{\citenamefont{Wiringa et~al.}(1995)\citenamefont{Wiringa, Stoks, and
  Schiavilla}}]{wiringa95}
\bibinfo{author}{\bibfnamefont{R.~B.} \bibnamefont{Wiringa}},
  \bibinfo{author}{\bibfnamefont{V.~G.~J.} \bibnamefont{Stoks}},
  \bibnamefont{and}
  \bibinfo{author}{\bibfnamefont{R.}~\bibnamefont{Schiavilla}},
  \bibinfo{journal}{Phys. Rev. C} \textbf{\bibinfo{volume}{51}},
  \bibinfo{pages}{38} (\bibinfo{year}{1995}).

\bibitem[{\citenamefont{Pudliner et~al.}(1995)\citenamefont{Pudliner,
  Pandharipande, Carlson, and Wiringa}}]{pudliner95}
\bibinfo{author}{\bibfnamefont{B.~S.} \bibnamefont{Pudliner}},
  \bibinfo{author}{\bibfnamefont{V.~R.} \bibnamefont{Pandharipande}},
  \bibinfo{author}{\bibfnamefont{J.}~\bibnamefont{Carlson}}, \bibnamefont{and}
  \bibinfo{author}{\bibfnamefont{R.~B.} \bibnamefont{Wiringa}},
  \bibinfo{journal}{Phys. Rev. Lett.} \textbf{\bibinfo{volume}{74}},
  \bibinfo{pages}{4396} (\bibinfo{year}{1995}).

\bibitem[{\citenamefont{Bajdich et~al.}(2006)\citenamefont{Bajdich, Mitas,
  Wagner, and Schmidt}}]{bajdich06b}
\bibinfo{author}{\bibfnamefont{M.}~\bibnamefont{Bajdich}},
  \bibinfo{author}{\bibfnamefont{L.}~\bibnamefont{Mitas}},
  \bibinfo{author}{\bibfnamefont{L.~K.} \bibnamefont{Wagner}},
  \bibnamefont{and} \bibinfo{author}{\bibfnamefont{K.~E.}
  \bibnamefont{Schmidt}} (\bibinfo{year}{2006}),
  \eprint{arXiv:cond-mat/0610850}.

\bibitem[{\citenamefont{Astrakharchik et~al.}(2004)\citenamefont{Astrakharchik,
  Boronat, Casulleras, and Giorgini}}]{astrakharchik04}
\bibinfo{author}{\bibfnamefont{G.~E.} \bibnamefont{Astrakharchik}},
  \bibinfo{author}{\bibfnamefont{J.}~\bibnamefont{Boronat}},
  \bibinfo{author}{\bibfnamefont{J.}~\bibnamefont{Casulleras}},
  \bibnamefont{and} \bibinfo{author}{\bibfnamefont{S.}~\bibnamefont{Giorgini}},
  \bibinfo{journal}{Phys. Rev. Lett.} \textbf{\bibinfo{volume}{93}},
  \bibinfo{pages}{200404} (\bibinfo{year}{2004}).

\bibitem[{\citenamefont{Carlson and Reddy}(2005)}]{carlson05}
\bibinfo{author}{\bibfnamefont{J.}~\bibnamefont{Carlson}} \bibnamefont{and}
  \bibinfo{author}{\bibfnamefont{S.}~\bibnamefont{Reddy}},
  \bibinfo{journal}{Phys. Rev. Lett.} \textbf{\bibinfo{volume}{95}},
  \bibinfo{pages}{060401} (\bibinfo{year}{2005}).

\end{thebibliography}
\end{document}